\documentclass[10pt,twocolumn]{article}

\usepackage[preprint]{antintlpaper}
\usepackage{float}
\usepackage{cuted}
\usepackage{algorithm}
\usepackage{algpseudocode}
\usepackage{simpleicons}
\newcommand{\benchmarkname}{BizAgentBench}
\newcommand{\numTasks}{50}

\AntTitle{BizAgentBench: Unveiling Hidden Challenges\texorpdfstring{\\}{ }in Deploying Long-Horizon Agents\texorpdfstring{\\}{ }for Real-World Business}
\AntRunningTitle{BizAgentBench: Long-Horizon Agents for Real-World Business}
\AntAuthors{%
  \texorpdfstring{\mbox{\href{https://openreview.net/profile?id=~Enhan_Zhao1}{Enhan Zhao}\textsuperscript{1,2,*}}}{Enhan Zhao} \and
  \texorpdfstring{\mbox{\href{https://openreview.net/profile?id=~Yongqi_Tong2}{Yongqi Tong}\textsuperscript{2,*}}}{Yongqi Tong} \and
  \texorpdfstring{\mbox{\href{https://openreview.net/profile?id=~Yu_Xuejun1}{Xuejun Yu}\textsuperscript{2}}}{Xuejun Yu} \and
  \texorpdfstring{\mbox{\href{https://openreview.net/profile?id=~Hang_Wang14}{Hang Wang}\textsuperscript{2}}}{Hang Wang} \and
  \texorpdfstring{\mbox{\href{https://openreview.net/profile?id=~Pan_Wang13}{Pan Wang}\textsuperscript{2}}}{Pan Wang}\texorpdfstring{\\[2pt]}{, }
  \texorpdfstring{\mbox{\href{https://openreview.net/profile?id=~yichao_wang7}{Yichao Wang}\textsuperscript{2}}}{Yichao Wang} \and
  \texorpdfstring{\mbox{\href{https://openreview.net/profile?id=~Di_He1}{Di He}\textsuperscript{1}}}{Di He} \and
  \texorpdfstring{\mbox{\href{https://openreview.net/profile?id=~JIANSHE_LI1}{Jianshe Li}\textsuperscript{2}}}{Jianshe Li} \and
  \texorpdfstring{\mbox{Jiang-Ming Yang\textsuperscript{2}}}{Jiang-Ming Yang} \and
  \texorpdfstring{\mbox{\href{https://openreview.net/profile?id=~Wei_Wu1}{Wei Wu}\textsuperscript{2,\ensuremath{\dagger}}}}{Wei Wu}%
}
\AntAffiliations{\textsuperscript{1}Peking University\quad\textsuperscript{2}Ant International}
\AntContact{Emails: \{zhaoenhan.zeh, tongyongqi.yq, xuejun.yu, wh549825, apen.wp, jmyang\}@ant-intl.com\\
\textsuperscript{\ensuremath{\dagger}}Corresponding author: Wei Wu (wuwei19850318@gmail.com).}
\AntFirstPageNote{\textsuperscript{*}\,Equal Contributions}
\AntDate{October 2026}
\AntLinks{%
  \AntResourceLink{https://github.com/ant-intl/BizAgentBench}{\textcolor[HTML]{24292F}{\simpleicon{github}}}{GitHub}\qquad
  \AntResourceLink[1.15em]{https://huggingface.co/datasets/ant-intl/BizAgentBench}{\includegraphics{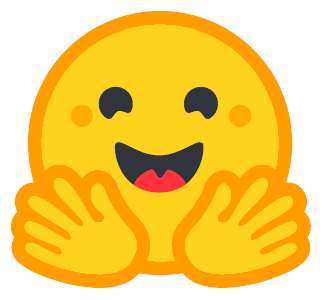}}{Hugging Face}%
}
\AntKeywords{long-horizon agents, business workflows, payment systems, benchmark evaluation}
\AntAbstract{
Existing long-horizon agent benchmarks often emphasize final task success while providing limited assessment of intermediate execution validity. Yet real-world business operations require agents to navigate extensive, potentially outdated or conflicting policies while satisfying critical constraints throughout execution. These requirements are particularly consequential in high-stakes domains such as payments, where intermediate violations can incur financial losses or regulatory consequences regardless of the final outcome. To bridge this evaluation gap, we introduce \benchmarkname{}, a multi-turn benchmark grounded in real-world procedures from a leading payments company. Its executable environment integrates 789 knowledge documents and 601 discoverable business tools, with consistent policies, states, and tool behavior. The benchmark comprises \numTasks{} tasks involving resource constraints, temporal dependencies, prior commitments, and conflicting or outdated guidance. Each task is validated through reference execution, while deterministic checks assess both goal attainment and execution validity without prescribing a single solution path. Experiments reveal substantial limitations even among frontier models: Claude Opus 5 and GPT-6 Sol achieve strict success rates of only 64.0\% and 61.3\%, respectively. For Qwen3.8-27B, goal attainment reaches 42.0\%, but strict success drops to 32.7\%, demonstrating how outcome-based evaluation can mask invalid execution. Beyond success rates, DeepSeek-V4.1-Flash averages 191.3 tool calls per run, approximately 70\% more than reference executions, revealing substantial execution overhead. Overall, \benchmarkname{} establishes a testbed for evaluating and advancing long-horizon agents beyond task completion, toward reliable and constraint-compliant execution in complex business environments.
}

\begin{document}
\makeanttitle

\section{Introduction}
\label{sec:introduction}

\begin{figure*}[!t]
    \centering
    \includegraphics[width=\textwidth]{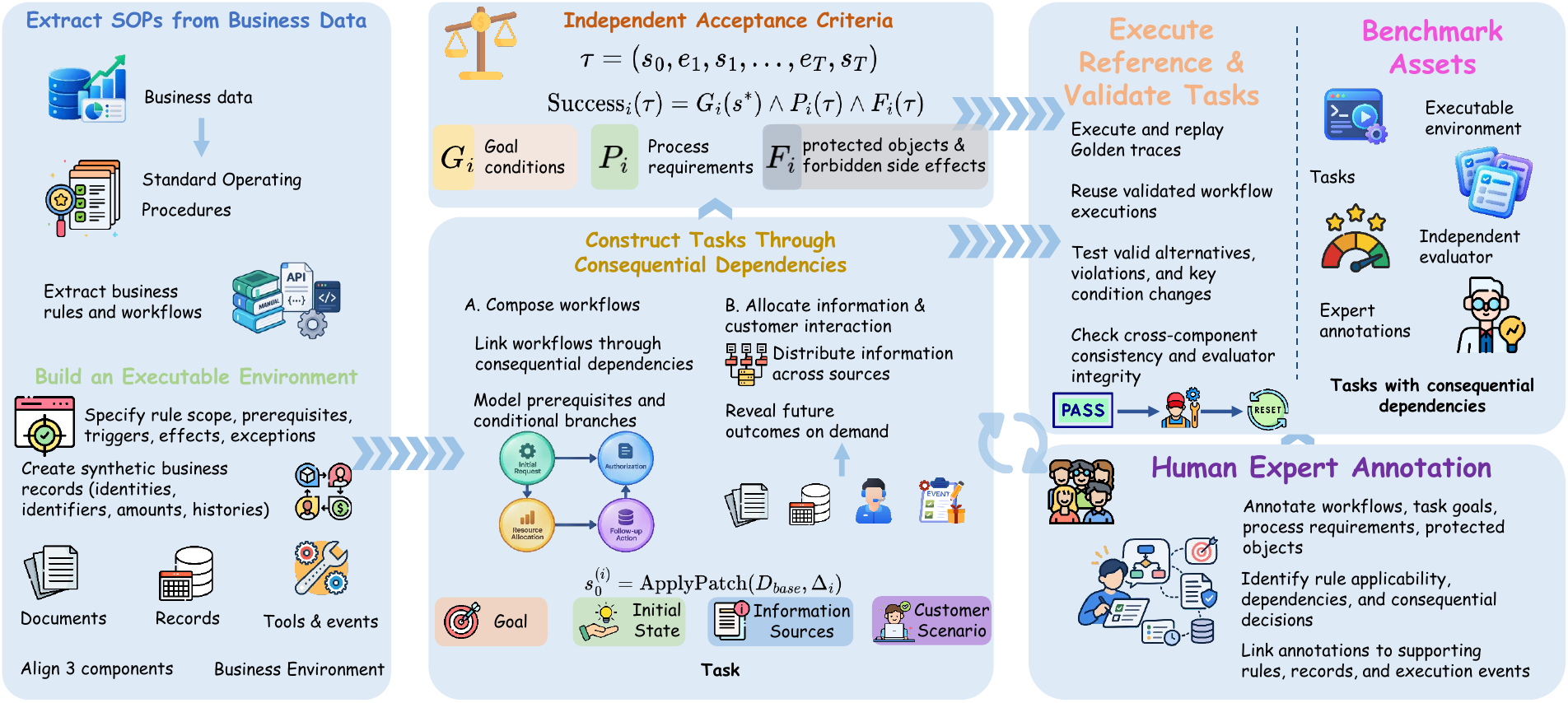}
    \caption{Overview of BizAgentBench construction. Business workflows are normalized into an executable environment; tasks combine consequential dependencies with separate goal, process, and protection criteria, and are validated using reference executions and targeted counterexamples.}
    \label{fig:benchmark-overview}
    \vspace{0pt}
\end{figure*}

Advances in artificial intelligence are transforming real-world business operations, with customer service providing a representative example. Early FAQ-based systems addressed user queries by matching them against curated question--answer pairs \citep{sakata_faq_retrieval,li_alime_assist}. Although effective for isolated queries, these systems offered limited support for contextual understanding across multiple turns. Subsequent advances in deep neural networks enabled end-to-end task-oriented dialogue systems, extending customer service from single-turn question answering to multi-turn task resolution \citep{raghu_flodial,budzianowski2018multiwoz,li2017end}. By tracking conversational context and incorporating external knowledge, these systems supported more sophisticated interactions, but remained largely constrained by predefined dialogue acts, slot schemas, and short-horizon state tracking. More recently, the emergence of large language models (LLMs) has driven a further transition toward agentic customer service, enabling increasingly autonomous execution of complex business workflows through policy interpretation, multi-step planning, and external tool orchestration \citep{yao_tau_bench,tong_onemodel}. Unlike conventional dialogue systems, such agents can take actions that alter business states, commit resources, and constrain subsequent decisions. Consequently, reliable deployment requires not only successful task completion but also adherence to business requirements throughout the execution process \citep{ruan_toolemu}.

This transition toward autonomous business execution introduces two fundamental challenges. First, agents must achieve the desired final outcome while maintaining execution validity at critical intermediate steps. While exploratory actions may be permissible, certain business constraints must remain satisfied throughout execution, and violating them can constitute task failure regardless of the eventual outcome. For example, an unauthorized payment remains a compliance violation even if the transferred funds are subsequently returned. Thus, successful task completion does not necessarily imply valid execution. Second, agents must identify and apply the authoritative business rules governing their actions. In real-world operations, business rules are distributed across heterogeneous sources, including policy manuals, FAQs, and operational notices \citep{shi_tau_knowledge}. These sources are maintained by different teams and updated at different times, allowing outdated guidance to coexist with current policies \citep{ouyang_hoh}. Moreover, seemingly conflicting instructions may differ in their applicability across jurisdictions, products, or effective periods. Consequently, retrieving relevant information is insufficient: agents must determine which rules are authoritative, current, and applicable to the specific business context \citep{nam_timelyrag}. Although existing long-horizon agent benchmarks evaluate policy-governed tool use and knowledge access \citep{lu_toolsandbox}, evaluations centered primarily on final task success may fail to capture violations of critical intermediate constraints or incorrect application of business rules. This limitation leaves an important gap between benchmark performance and the assurance required for reliable deployment in real-world business environments.

We therefore introduce BizAgentBench to evaluate whether agents can identify applicable rules before taking consequential actions and complete business requests while maintaining valid execution throughout. The benchmark comprises \numTasks{} long-horizon tasks grounded in real-world business workflows from a leading global payments company. These tasks combine tool use with multi-turn customer dialogue: a user simulator provides customer-specific information and responds to approval requests throughout execution. To test rule interpretation before action, we provide current and historical guidance with explicit scopes and effective periods across policy documents. Agents must determine which instructions govern the current request and use accompanying tool documentation to discover and invoke relevant tools. We instantiate these business workflows in a stateful environment, linking them through shared resources, timing requirements, and prior commitments. The environment retains action consequences, so later decisions must account for commitments already made.

To detect violations that final task success can conceal, a deterministic evaluator checks both outcomes and recorded execution histories against task goals and applicable rules. Success requires completing the requested goals while satisfying prior authorization requirements, persistent constraints, and restrictions on side effects. Successful task completion therefore cannot compensate for intermediate violations, while different legitimate strategies remain acceptable. Reference replays verify that goals and constraints are jointly satisfiable; alternative valid executions and targeted violations test whether the evaluator accepts valid solutions and rejects invalid ones.

Our experiments reveal that even advanced models such as GPT-6 Sol struggle on \benchmarkname{}, achieving a strict success rate of only 61.3\%. Performance also varies substantially across runs, with only 40.0\% of tasks successfully completed in all three attempts, raising concerns about the reliability of frontier models in real-world business deployments. By evaluating both final outcomes and execution histories, \benchmarkname{} identifies invalid executions that outcome-only evaluation would incorrectly deem successful. For instance, four DeepSeek-V4.1-Flash runs achieve their goals and finish above the required cash-reserve threshold, despite having breached it at intermediate steps. The challenges are even more pronounced for smaller models. Qwen3.5-9B, for example, averages 187.2 tool calls per run, more than half of which result in errors, primarily from attempts to access non-public resources. None of its 150 runs successfully achieves all task objectives. Together, these findings reveal complementary limitations in execution consistency, process compliance, and effective tool use.

\benchmarkname{} combines: (1) long-horizon tasks grounded in real-world payment workflows, linked by shared resources, deadlines, and commitments; (2) distributed guidance that requires identifying applicable rules across scopes, versions, and effective periods; and (3) deterministic checks of goals, process compliance, and protection requirements over business states and execution histories, while accepting different valid strategies. These characteristics establish \benchmarkname{} as a rigorous, business-grounded testbed for investigating how agents coordinate information gathering, planning, and action to accomplish complex business tasks while maintaining execution validity throughout.

\section{Related Work}
\label{sec:related_work}

\noindent\textbf{Long-horizon agent benchmarks.}
Agent benchmarks assess reasoning and tool use in interactive environments \citep{liu_agentbench,mialon_gaia}.
Web and desktop benchmarks cover shopping, browsing, and computer operation \citep{yao_webshop,deng_mind2web,zhou_webarena,koh_visualwebarena,xie_osworld}.
SWE-bench evaluates repository-level issue resolution, while Terminal-Bench tests broader technical tasks through final-state checks \citep{jimenez_swe_bench,merrill_terminal_bench}.
SWE-Milestone checks feature completion and regressions at successive milestones \citep{deng_swe_milestone}; Long-Horizon-Terminal-Bench measures partial progress through subtask checks on the final state \citep{li_long_horizon_terminal_bench}.
While these benchmarks emphasize technical progress and software integrity, \benchmarkname{} models business environments with scattered tool documentation, noisy, outdated, or conflicting information, and changing conditions.
It checks intermediate actions against applicable business requirements.
An action's validity depends on the approvals, funds, and policy in force when it occurs.

\noindent\textbf{Agents in business workflows.}
Business-oriented benchmarks introduce operational rules and organizational context.
The $\tau$ family studies policy-governed interaction, user--agent coordination, and knowledge-grounded workflows \citep{yao_tau_bench,barres_tau2_bench,shi_tau_knowledge}.
Enterprise suites cover browser-based knowledge work, compositional workflows, workplace collaboration, and CRM operations \citep{drouin_workarena,boisvert_workarena_plus,xu_theagentcompany,huang_crmarena,huang_crmarena_pro}.
AutomationBench evaluates final application states without prescribing tool-call sequences, preserving flexibility in how agents solve tasks \citep{shepard_automationbench}.
\benchmarkname{} likewise allows exploration and alternative strategies, while checking authorization and other business requirements at consequential steps, such as whether a transfer was approved before execution.

\noindent\textbf{Long-horizon planning and adaptation.}
Tool-use learning and reasoning--action interleaving provide foundations for language agents \citep{schick_toolformer,yao_react}.
Reflection, skill reuse, and search support multistep reasoning and planning \citep{shinn_reflexion,wang_voyager,yao_tree_of_thoughts,zhou_lats}, while clarification research examines when additional information is useful \citep{gulati_clarification_timing}.
\benchmarkname{} focuses on replanning in business systems where permitted actions can undermine later obligations and both acting and waiting can incur costs.
Plans must accommodate commitments already made.

\section{\benchmarkname{}}
\label{sec:bizagentbench}

\benchmarkname{} evaluates agents that execute customer requests across real-world business workflows.
Each task combines an objective, an initial business state, a customer scenario, and acceptance criteria.
Agents use a shared environment of documents, records, and executable tools.
Success requires fulfilling the request while respecting the requirements that govern its intermediate execution.

\subsection{Business Context and Reconstructed Environment}
\label{subsec:business_environment}

\benchmarkname{} draws on business workflows from one of the world's leading payments and financial technology companies.
The company supports merchants and enterprises through cross-border payments, foreign exchange, and treasury services.
Customer requests can span multiple currencies, legal entities, and financial operations.
In practice, deployed agents must reconcile fragmented records and apply the right rules, even when payments or reviews are still pending.
An action accepted by one system can still consume funds or capacity needed to fulfill a later obligation.

To preserve the confidentiality of business rules, we abstract sensitive details from the source workflows.
We reconstruct their key business logic and execution mechanisms in an executable environment.
Agents consult documents for business rules and tool documentation, and query records to establish the customer's current business state.
They interact with a user simulator to obtain private customer information and required approvals.
Business tools allow agents to execute operations and inspect their results.
The frozen release contains \numTasks{} tasks and 789 knowledge documents.
Agents have nine always-available tools through which they discover and invoke 601 business capabilities; the user simulator has ten tools.
Each run uses an independent copy of the database.
Appendix~\ref{sec:statistics_details} reports the environment inventory and Golden trace statistics.

\subsection{Construction Insights from Our Business Workflows}
\label{subsec:construction_insights}

\noindent\textbf{Outdated and conflicting guidance.}
Business documents can remain accessible after their instructions have been superseded.
Apparent conflicts can also arise because two rules govern different products, regions, or effective periods.
Resolving a conflict requires establishing which source is authoritative and which rule applies to the particular business event.
An expired insurance policy, for example, may still cover a past incident.
\benchmarkname{} includes current and historical guidance with explicit scopes and effective periods, making rule selection part of executing the request.

\noindent\textbf{Authorization and execution validity.}
General consent does not substitute for approval of a specific operation.
Approval must cover the actual operation and its terms, come from an authorized approver, and precede execution.
Other requirements, such as a minimum reserve or continuous guarantee coverage, must hold throughout the workflow.
An operation accepted by a tool can still breach these requirements.
For example, an account holds \$60 and must retain at least \$20 throughout execution.
Paying a \$60 retainer leaves the account at zero and immediately breaches this requirement, even though the payment is accepted.
Later refunds and adjustments can restore the closing reserve without repairing the earlier breach (Appendix~\ref{sec:goal_process_cases}).
\benchmarkname{} therefore evaluates the relevant approvals and state changes throughout execution alongside the requested outcomes.

\noindent\textbf{Delayed information and time-sensitive decisions.}
Reviews, refunds, and settlements may finish after another obligation requires action.
In the delivery case (Appendix~\ref{case:delivery_window}), the priority booking window closes before warehouse confirmation arrives.
Waiting for that confirmation makes the first delivery deadline unattainable.
\benchmarkname{} represents when information becomes observable and when actions become available, so that waiting can change whether the request remains feasible.

\subsection{Evaluating Outcomes and Execution}
\label{subsec:acceptance_criteria}

Let $\tau=(s_0,e_1,s_1,\ldots,e_T,s_T)$ denote an execution history, where $e_t$ is a tool execution or environment event.
Task $i$ succeeds when
\begin{equation}
    \operatorname{Success}_i(\tau)
    = G_i(\tau) \land P_i(\tau) \land F_i(\tau).
    \label{eq:construction_success}
\end{equation}
Here, $G_i$ checks the requested outcomes and permitted terminal conditions, $P_i$ checks requirements governing execution, and $F_i$ checks protected objects and prohibited side effects.
These categories are defined by the business requirements being checked; each uses the relevant business states and execution histories as evidence.
For example, a payment goal may require an actual settlement receipt and its timestamp, while a reserve requirement needs the balance after each relevant state change.

Checks operate on business states, ledgers, receipts, and event histories.
They distinguish an executed action from a proposal, an approval, or a rejected call without side effects.
Different valid action sequences can satisfy the same criteria; the evaluator does not match executions to a Golden trace.
The primary score is computed by deterministic, executable checks over business states and execution histories, without using an LLM judge.

\section{Data Construction}
\label{sec:construction}

We construct multi-turn tasks in which the agent and a user simulator work together to achieve a shared business goal.
Drawing on the insights in Section~\ref{subsec:construction_insights}, we retain current and historical guidance with explicit applicability conditions, introduce temporal dependencies between workflows, and check authorization and intermediate execution against business requirements.
We organize data construction into five stages.
First, we organize business rules into SOPs and adapt them for simulation (Section~\ref{subsec:sop_extraction}).
Second, we build executable tasks by defining states, operations, and acceptance checks, and linking workflows through resource and timing constraints (Section~\ref{subsec:executable_environment}).
Third, we configure a user simulator to disclose customer information and perform customer-side actions through multi-turn interaction (Section~\ref{subsec:user_simulator_setup}).
Fourth, we construct and replay Golden traces to establish executable strategies that satisfy the task requirements (Section~\ref{subsec:construction_validation}).
Finally, we validate task behavior and acceptance checks using alternative executions, violations, and variations in task conditions and execution, alongside expert review (Section~\ref{subsec:human_expert_annotation}).

\subsection{Grounding Tasks in Business Workflows}
\label{subsec:sop_extraction}

We organize source business documents into structured SOPs that capture the rules and dependencies needed for executable tasks.
As summarized in Algorithm~\ref{alg:sop-construction}, we extract candidate workflows with supporting source evidence and group related candidates into workflow plans that organize business branches.
These plans are converted into SOPs specifying steps, decisions, transitions, and return paths.
The associated rules record their scope, affected objects, prerequisites, actions, and exceptions.

\begin{algorithm*}[!t]
    \caption{Constructing SOPs from business documents}
    \label{alg:sop-construction}
    \small
    \algrenewcommand{\algorithmicrequire}{\textbf{Input:}}
    \algrenewcommand{\algorithmicensure}{\textbf{Output:}}
    \begin{algorithmic}[1]
        \Require Business documents $\mathcal D$
        \Ensure SOPs $\widetilde{\mathcal S}$
        \State $\mathcal C \gets$ extract workflow candidates from $\mathcal D$, retaining source evidence links
        \State $\mathcal P \gets$ group related workflow candidates into hierarchical workflow plans with evidence links
        \State $\mathcal S \gets$ generate SOPs from $\mathcal P$ and their source evidence
        \State $\widetilde{\mathcal S} \gets \emptyset$
        \For{each SOP $s \in \mathcal S$}
            \State Specify steps, decisions, next-step links, and return links in $s$
            \State Review $s$ against its source evidence via the workflow plan and candidates
            \State Exclude content with unresolved conflicts
            \State Attach reviewed knowledge to $s$ and its steps, retaining applicability conditions and source evidence
            \State $\tilde{s} \gets$ rewrite sensitive business content in $s$ for simulation
            \State $\widetilde{\mathcal S} \gets \widetilde{\mathcal S} \cup \{\tilde{s}\}$
        \EndFor
        \State \Return $\widetilde{\mathcal S}$
    \end{algorithmic}
\end{algorithm*}
\vspace{0pt}

We review the extracted rules against their source evidence and exclude content whose meaning or applicability cannot be established.
Differences that can be resolved through scope, version, or effective period are preserved and carried forward into the environment's documentation.
This allows current and historical guidance to coexist while providing the evidence needed to determine which instruction governs a particular operation.

To protect proprietary information, we rewrite sensitive business details consistently across related rules while preserving their decision logic and dependencies.
We distinguish requirements grounded in the source documents, interpretations needed to make them executable, and additional constraints introduced during task construction.
The resulting SOPs provide the business requirements used to construct the environment and tasks.

\subsection{Building Executable Tasks}
\label{subsec:executable_environment}
\label{subsec:task_composition}

From these SOPs, we construct an executable environment with aligned documents, records, and tools.
Documents specify business rules and tool interfaces; synthetic records instantiate entities, accounts, obligations, authorizations, and transaction histories; tools and scheduled events implement the corresponding state changes.
For each operation, we align its documented conditions, relevant records, and implemented effects.
To represent outdated and conflicting guidance, the synthesized documents include current and historical rules and guidance for related products, with explicit scopes, versions, and effective periods.

We construct tasks by selecting workflows that share business objects or resources, or whose outputs establish prerequisites for other operations.
Each task specifies a customer objective, an initial database state, and scheduled events.
Additional constraints introduced during task construction are stated in public documents or customer requirements.
The facts needed to apply these constraints are available through record queries or customer interaction.

We connect workflows through shared resources, state prerequisites, and timing requirements.
Temporal dependencies link scheduled events in one workflow to actions and deadlines in another.
Together with delayed information availability, these dependencies create the time-sensitive decisions described in Section~\ref{subsec:construction_insights}: waiting for an external result may close a required execution window.
Because workflows share funds or authorization limits, a permitted action in one can consume resources needed by another obligation.
Agents must therefore coordinate decisions across workflows and account for earlier actions when revising their plans.
Dependency graphs capture these relationships and conditional branches during task construction.

Acceptance criteria are derived from the customer objective and applicable rules, jointly with the task's initial conditions and dependencies, before generating Golden traces.
Requested outcomes and permitted terminal conditions determine the goal checks $G$; requirements governing execution determine the process checks $P$; and protected objects and prohibited side effects determine the protection checks $F$.
We link these checks to their governing rules or customer requirements and identify the relevant evidence in records, ledgers, receipts, and execution histories.
For authorization and execution validity, intermediate checks verify that required approval covers the executed terms and precedes the operation, and test continuous constraints at each relevant state change.
The checks evaluate the specified business conditions and allow different valid action sequences.
Golden traces subsequently establish that the requirements are jointly satisfiable, as described in Section~\ref{subsec:construction_validation}.

\subsection{Setting Up the User Simulator}
\label{subsec:user_simulator_setup}
\label{subsec:information_allocation}

We configure an LLM as the user simulator for each task using a common interaction protocol and a task-specific customer scenario.
The scenario specifies the customer's objective, private facts, preferences, supporting materials, and opening request.
The common protocol defines the customer role, permitted customer-side actions, and turn-taking behavior.
Together, they guide the simulator's responses based on the conversation and the results of its own tool calls.

The interaction begins with the opening request; other customer information is disclosed when relevant questions or proposals arise.
The agent must ask for the constraints and materials needed to proceed, while the simulator answers from its scenario and customer-facing receipts.
Fixed facts and constraints remain consistent across turns.
Where the scenario permits negotiation, a revised preference must be stated explicitly.
This requires the agent to elicit customer-specific requirements and retain their terms as the business workflow progresses.

The simulator also performs customer-side operations in the shared environment, such as submitting an available document or approving a proposed transaction.
Approval is recorded against a specific request and its current terms; changed terms require renewed approval.
Factual questions take priority over action requests.
The simulator performs one requested write or wait at a time, reports its actual result, and returns control to the agent.
If a requested action fails or is unclear, it reports the problem and asks for clarification.
The resulting dialogue alternates customer input and actions with the agent's investigation and execution until the interaction ends.

\subsection{Constructing and Replaying Golden Traces}
\label{subsec:construction_validation}

We construct Golden traces to establish that each task has at least one executable strategy satisfying its goal, process, and protection requirements.
This is especially important for linked workflows, where shared resources and timing constraints can make otherwise feasible workflows incompatible.

We construct reference strategies from the SOPs, task conditions, and dependencies between workflows.
Each strategy is executed through the tools and customer-side actions used in evaluation.
The resulting Golden trace records the actions and their consequences.

We replay each trace from the task's initial state and evaluate it using the goal, process, and protection checks defined for the task (Section~\ref{subsec:executable_environment}).
Even when individual workflows have already been validated, we replay the complete task to check their interactions through shared resources, state prerequisites, and timing constraints.
Each of the \numTasks{} released tasks has a passing Golden trace.

\subsection{Task Validation and Expert Review}
\label{subsec:human_expert_annotation}

Task validation checks whether the evaluator accepts valid executions and detects violations of the stated requirements.
We test valid alternatives that use different permitted routes or action orders, and targeted violation traces that break specific requirements.
A temporary reserve breach, for example, must fail the process check even if later transactions restore the balance.

We also change facts that determine whether an operation is valid or feasible, and verify that the same acceptance checks yield the expected judgments under the modified conditions.
Such changes may require a different execution strategy.
In contrast, operational variations that preserve compliance and business effects should leave the evaluation outcome unchanged.
In particular, an idempotent retry permitted by the tool contract should produce no additional business effect.

Five researchers and a team of business-domain experts reviewed the tasks and acceptance checks against customer requests, documents, business records, and validation traces.
They examined whether the linked workflows formed coherent business scenarios, whether agents could obtain the information needed for their decisions, and whether each check represented its governing rule or customer requirement.
Identified inconsistencies in the documents, tools, simulator, or evaluator were corrected, and the affected executions were rerun.
The final benchmark includes only tasks that passed both validation and expert review.

\section{Experiments}
\label{sec:experiments}

Our experiments examine whether agents complete customer requests while satisfying the requirements that govern execution.
We report goal, process, protection, and strict success across models, then use information ablations and execution traces to study customer interaction, document retrieval, and long-horizon tool use.
We also assess the stability of the user simulator.

\subsection{Experimental Setup}
\label{subsec:experimental_setup}

We evaluate \benchmarkname{} on the same \numTasks{} tasks with proprietary and open-source models \citep{openai2026gpt6astra,deepseekai2026deepseekv41flashpushinglimitskv,deepseekai2026deepseekv4highlyefficientmilliontoken,yang2025qwen3technicalreport} (Table~\ref{tab:gpf_results}).
Each run starts from an independent copy of the initial database and uses the same public documents, tools, and acceptance criteria.
In the default condition, agents retrieve documents, discover business capabilities, and interact with a DeepSeek-V4.1-Flash user simulator.
Golden traces, executable checks, and unrevealed event outcomes remain unavailable to the agent.

\begin{table*}[!t]
    \centering
    \small
    \caption{Goal ($G$), process ($P$), protection ($F$), intermediate failure, and strict success (\%). Intermediate failure means that a process or protection requirement failed. Claude Opus 5 and Muse Spark 1.3 have one run per task; the other models have three.}
    \label{tab:gpf_results}
    \setlength{\tabcolsep}{2.5pt}
    \renewcommand{\arraystretch}{1.08}
    \begin{tabular*}{\linewidth}{@{\extracolsep{\fill}}lrrrrr@{}}
        \toprule
        \textbf{Model}
        & \textbf{Goal} $\mathbf{(G)}$
        & \textbf{Process} $\mathbf{(P)}$
        & \textbf{Protection} $\mathbf{(F)}$
        & \textbf{IF} $\mathbf{(\neg(P\land F))}$
        & \textbf{Strict} $\mathbf{(G\land P\land F)}$ \\
        \midrule
        Claude Opus 5          & 64.0 & 96.0 &  98.0 &  4.0 & \textbf{64.0} \\
        GPT-6 Sol              & 64.7 & 94.7 & 100.0 &  5.3 & \textbf{61.3} \\
        GLM-5.3-Flash          & 48.0 & 79.3 &  87.3 & 29.3 & \textbf{43.3} \\
        DeepSeek-V4.1-Flash    & 46.7 & 92.0 &  96.7 &  9.3 & \textbf{43.3} \\
        DeepSeek-V4-Pro        & 41.3 & 85.3 &  88.7 & 24.7 & \textbf{39.3} \\
        Muse Spark 1.3         & 38.0 & 94.0 &  98.0 &  6.0 & \textbf{38.0} \\
        Qwen3.8-27B            & 42.0 & 74.0 &  93.3 & 28.0 & \textbf{32.7} \\
        DeepSeek-V4-Flash      &  7.3 & 53.3 &  95.3 & 47.3 & \textbf{5.3} \\
        Qwen3.5-397B-A17B      &  1.3 & 46.7 &  93.3 & 56.7 & \textbf{0.7} \\
        Qwen3.5-9B             &  0.0 & 18.7 &  89.3 & 83.3 & \textbf{0.0} \\
        \bottomrule
    \end{tabular*}
    \vspace{0pt}
\end{table*}

\begin{figure*}[!t]
    \centering
    \includegraphics[width=\textwidth]{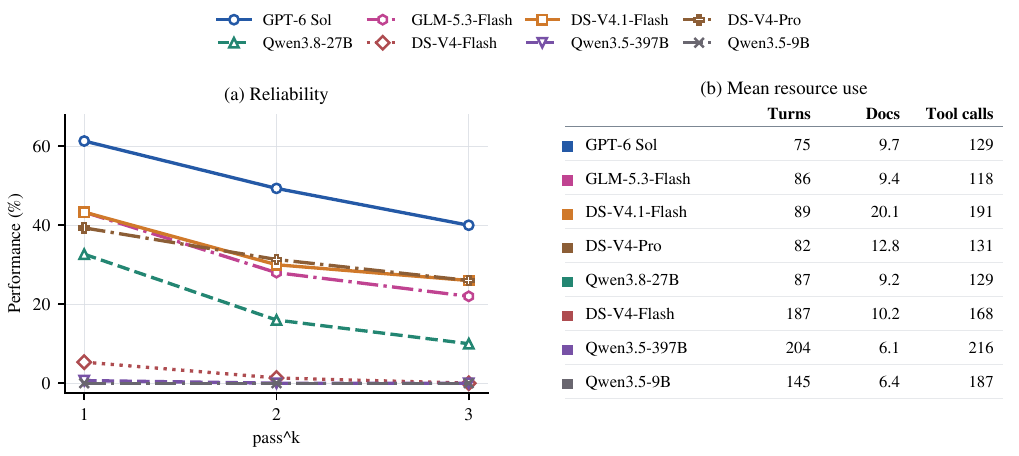}
    \caption{Reliability and resource use for models with three runs per task. Left: pass\textasciicircum{}k averages, over all $k$-run subsets, the fraction of tasks that succeed in every selected run. Right: mean assistant turns, distinct documents read, and assistant tool calls per run.}
    \label{fig:overall_performance}
    \vspace{0pt}
\end{figure*}

The runner imposes no cumulative limit on generated tokens, tool calls, or scheduler steps, and sets per-response output-token limits high enough to avoid routine truncation.
Runs retain a wall-clock timeout and a tool-error threshold.
Each model request includes the accumulated dialogue and tool history.
Following \citet{shi_tau_knowledge}, if a provider reports context-window overflow, the oldest quarter of eligible knowledge-search and document-read tool responses is replaced in the model-facing context, while the complete execution record is preserved.

For models evaluated three times per task, we average single-run success across the three runs; other models have one run per task.
Information ablations use the same tasks, bank-side tools, database, and acceptance criteria.
We track infrastructure and simulator protocol failures separately from business outcomes, and resume transient transport interruptions from the verified prior state.

\subsection{Overall Performance and Process Validity}
\label{subsec:overall_process_validity}

The main comparison separates goal attainment from valid execution using the criteria in Equation~\ref{eq:construction_success}.
Table~\ref{tab:gpf_results} reports the pass rates for goals and permitted terminal conditions ($G$), process requirements ($P$), protection requirements ($F$), strict success ($G\land P\land F$), and intermediate failure (IF, $\neg(P\land F)$).
We report the predefined goal, process, and protection checks separately, with strict success requiring all three.

Frontier models perform well but leave substantial room for improvement: Claude Opus 5 achieves the highest strict success rate at 64.0\%, followed by GPT-6 Sol at 61.3\%.
GLM-5.3-Flash and DeepSeek-V4.1-Flash each achieve 43.3\% strict success, followed by DeepSeek-V4-Pro at 39.3\%, while Qwen3.5-9B completes none of its 150 runs.
Every model incurs intermediate failures, from 2 of 50 runs (4.0\%) for Claude Opus 5 to 125 of 150 (83.3\%) for Qwen3.5-9B; such failures are generally more frequent among lower-scoring models.
Process pass rates are below protection pass rates for every model, spanning 18.7--96.0\% for $P$ versus 87.3--100.0\% for $F$.
Figure~\ref{fig:overall_performance}(a) shows declining reliability: at $k=3$, GPT-6 Sol reaches 40.0\%, DeepSeek-V4.1-Flash 26.0\%, and GLM-5.3-Flash 22.0\%.
At $k=3$, Qwen3.8-27B reaches 10.0\%, and DeepSeek-V4-Flash and both Qwen3.5 models reach 0\%.
These declines indicate that the models remain unreliable across repeated runs.

\subsection{Information Acquisition}
\label{subsec:knowledge_retrieval}

\begin{table*}[!t]
    \centering
    \begin{minipage}[t]{0.47\linewidth}
        \centering
        \vspace{0pt}
        \caption{DeepSeek-V4.1-Flash strict success under three information conditions on 50 tasks.}
        \label{tab:information_ablation}
        \setlength{\tabcolsep}{4pt}
        \renewcommand{\arraystretch}{1.08}
        \begin{tabular}{@{}p{0.72\linewidth}r@{}}
            \toprule
            \textbf{Information condition} & \textbf{Strict (\%)} \\
            \midrule
            Default condition & 43.3 \\
            + Gold-doc & 50.0 \\
            + Full customer disclosure & 70.0 \\
            \bottomrule
        \end{tabular}
    \end{minipage}\hfill
    \begin{minipage}[t]{0.49\linewidth}
        \centering
        \vspace{0pt}
        \caption{DeepSeek-V4.1-Flash strict success with tool-call caps (three-run mean); $R_i$ is the Golden call count.}
        \label{tab:tool_budget}
        \setlength{\tabcolsep}{8pt}
        \renewcommand{\arraystretch}{1.08}
        \begin{tabular}{@{}lr@{}}
            \toprule
            \textbf{Tool-call cap} & \textbf{Strict (\%)} \\
            \midrule
            $1.2\times R_i$ & 0.7 \\
            $1.5\times R_i$ & 6.7 \\
            $2.0\times R_i$ & 26.7 \\
            No cap & 43.3 \\
            \bottomrule
        \end{tabular}
    \end{minipage}
    \vspace{0pt}
\end{table*}

We examine how information acquisition affects task performance, focusing on two sources: business documents and customer interaction.
Gold-doc provides applicable documents directly, while full customer disclosure supplies all customer-known task information in the opening request without a user simulator.
Required customer approvals remain in force.

Table~\ref{tab:information_ablation} compares these conditions with the default condition.
Strict success increases from 43.3\% to 50.0\% with Gold-doc and to 70.0\% with full customer disclosure.
The larger gain from full disclosure implies that eliciting customer-specific constraints and retaining their terms through execution are particularly difficult.
Gold-doc removes the search for applicable documents, but agents must still interpret their rules and apply them across linked actions.
Its smaller gain suggests that document retrieval accounts for only part of the information-use challenge.
Even with all customer facts available at the outset, 30.0\% of tasks remain unsolved, highlighting the remaining difficulty of planning and valid execution.

\subsection{Long-Horizon Behavior and Budget Sensitivity}
\label{subsec:horizon_budget}

The Golden traces average 111.8 assistant tool calls. DeepSeek-V4.1-Flash averages 191.3 calls (1.71 times the reference).
The excess is concentrated in information access: database queries average 111.1 versus 59.2 in the Golden traces, and knowledge searches or document reads average 43.3 versus 11.5, while business capability calls and request executions are similar (26.9 versus 28.7).

Figure~\ref{fig:overall_performance}(b) shows that longer traces need not reflect greater progress across models.
GLM-5.3-Flash matches DeepSeek-V4.1-Flash in strict success with 118.3 versus 191.3 tool calls and 9.4 versus 20.1 documents; Qwen3.5-397B-A17B uses 216.4 calls but succeeds only once in 150 runs.
Appendix~\ref{sec:execution_analysis} reports the full call distributions.

Table~\ref{tab:tool_budget} retrospectively limits each task to $\lceil\alpha R_i\rceil$ assistant tool calls, where $R_i$ is its Golden trace call count.
Across three runs per task, strict success falls from 65/150 (43.3\%) without a cap to 1/150 (0.7\%) at $1.2\times$, 10/150 (6.7\%) at $1.5\times$, and 40/150 (26.7\%) at $2.0\times$.
Even twice the reference call count excludes many otherwise successful runs, quantifying the cost of the agents' information search and repeated checking.

\subsection{Trajectory Analysis}
\label{subsec:trajectory_analysis}

\begin{table*}[!t]
    \centering
    \small
    \caption{Execution-trace statistics across stronger and weaker models. Counts are means per run; repeated database reads have identical tool names, arguments, and responses. The last column counts runs reaching the goal but failing process or protection requirements.}
    \label{tab:trajectory_analysis}
    \setlength{\tabcolsep}{2.5pt}
    \renewcommand{\arraystretch}{1.08}
    \begin{tabular*}{\linewidth}{@{\extracolsep{\fill}}lrrrr@{}}
        \toprule
        \textbf{Model} & \textbf{Assistant tool calls} & \textbf{Repeated database reads} & \textbf{Tool errors} & $\mathbf{G\land\neg(P\land F)}$ \\
        \midrule
        DeepSeek-V4.1-Flash & 191.3 & 16.4 & 2.0 & 5/150 \\
        Qwen3.8-27B         & 129.2 &  7.8 & 4.2 & 14/150 \\
        DeepSeek-V4-Flash   & 168.0 & 23.7 & 8.2 & 3/150 \\
        Qwen3.5-9B          & 187.2 & 16.5 & 101.6 & 0/150 \\
        \bottomrule
    \end{tabular*}
    \vspace{0pt}
\end{table*}

\noindent\textbf{Progress and repetition.}
Table~\ref{tab:trajectory_analysis} distinguishes additional tool use from literal repetition: a repeated database call has the same tool, arguments, and returned content as an earlier call in that run.
DeepSeek-V4.1-Flash makes 16.4 such calls per run; DeepSeek-V4-Flash makes 23.7 despite making fewer tool calls overall.
Qwen3.5-9B receives 101.6 tool-error replies per run, 71\% of them for querying non-public resources, and reads only 6.4 distinct documents, compared with 20.1 for DeepSeek-V4.1-Flash.
Thus, long traces mix information gathering with redundant reads and invalid navigation; the number of tool calls alone does not measure progress.

\noindent\textbf{Intermediate violations.}
Among DeepSeek-V4.1-Flash runs, five reach the requested goal but fail the full criteria: four cross a required cash-reserve floor before later restoring the balance, and one attempts execution before approval and receives a rejection.
Qwen3.8-27B has 14 such runs, including eight rejected pre-approval execution attempts; DeepSeek-V4-Flash has three.
Qwen3.5-9B fails process requirements in 122/150 runs, including 88 pre-approval execution attempts, but reaches no goals.
These histories explain why final goal checks alone overstate valid execution.

\noindent\textbf{Business-document scope.}
Golden traces contain reads of 5.4 distinct documents per task, whereas DeepSeek-V4.1-Flash opens 20.1.
Additional reads include documents on related payment operations and products with similar terminology but different applicability.
This shows that the complexity of the knowledge base makes identifying applicable guidance challenging for the model.
Appendix~\ref{sec:business_document_scope_example} gives a tax-payment example.

\subsection{Simulator Reliability and Evaluation Robustness}
\label{subsec:simulator_robustness}

We assess whether the user simulator follows the interaction protocol under two prompt variants.
The revised prompt explicitly specifies the order of customer actions and when to return control to the agent.
Using DeepSeek-V4.1-Flash for both roles, we observe protocol violations on 4 of 50 tasks with the original prompt and none with the revised prompt.
These violations involve batched approvals or further tool use before returning control.
We therefore adopt the revised prompt as the default user-simulator prompt in our evaluations.
Appendix~\ref{sec:simulator_patch_ablation} details the protocol and paired analysis.

\begin{figure}[!htbp]
    \centering
    \includegraphics[width=0.88\linewidth]{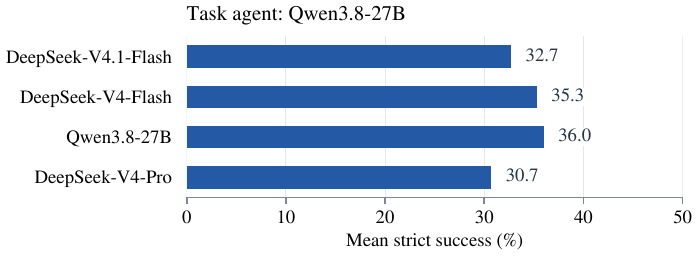}
    \caption{User-simulator model comparison using Qwen3.8-27B as the agent. Bars show the mean strict success rate over three runs; DeepSeek-V4.1-Flash is the default simulator.}
    \label{fig:simulator_model_sensitivity}
    \vspace{0pt}
\end{figure}

We also examine sensitivity to the choice of user-simulator model.
Qwen3.8-27B executes the tasks in all four conditions.
Figure~\ref{fig:simulator_model_sensitivity} reports the mean strict success rate over three runs for each simulator.
The mean rates range from 30.7\% to 36.0\%.
The largest deviation from the default DeepSeek-V4.1-Flash setting (32.7\%) is 3.3 percentage points.
This small spread suggests that agent performance is relatively robust to the choice of user simulator in this comparison.

\section{Conclusion}
\label{sec:conclusion}

With \benchmarkname{}, business requirements become testable conditions on both the outcomes and execution of long-horizon tasks. Our experiments show that task completion, process compliance, and consistency across runs reveal different weaknesses in current agents. Information ablations suggest that obtaining the right customer facts can substantially improve performance, while trajectory analysis shows that extensive tool use does not necessarily indicate meaningful progress. These findings motivate closer coordination between information gathering and action selection: agents need to establish when they have sufficient evidence to proceed and which obligations must remain satisfied as execution unfolds.

\bibliographystyle{plainnat}

\clearpage
\appendix
\begin{strip}
    \begin{minipage}{\textwidth}
    \captionsetup{type=figure}
    \centering
    \includegraphics[width=\textwidth]{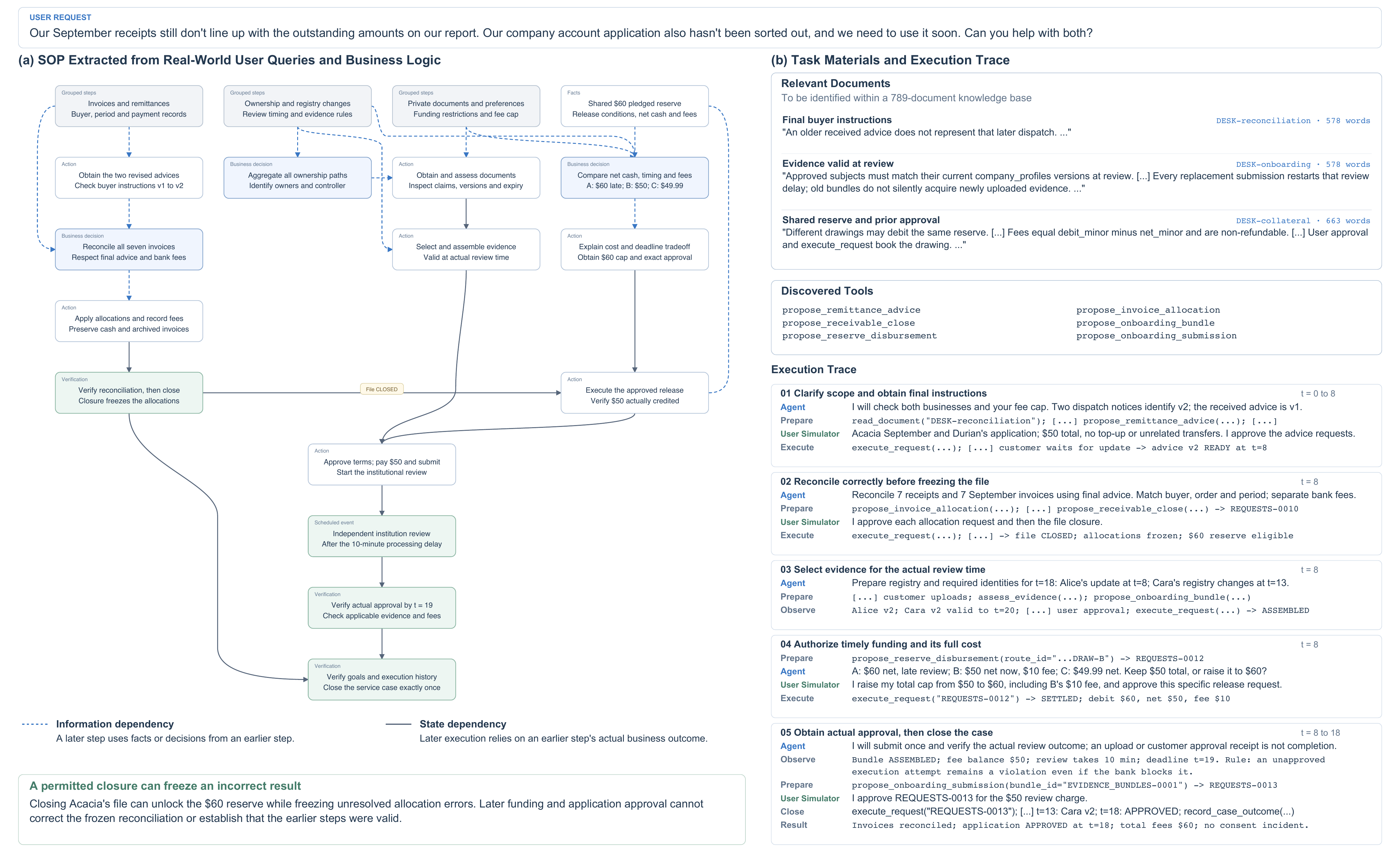}
    \caption{\textbf{A complete task example in \benchmarkname{}.}
    In Q03, the customer asks to reconcile Acacia's September receipts and complete Durian Trading's account application.
    \textbf{(a)} The SOP connects two rule-applicability decisions---resolving superseded buyer instructions and selecting evidence valid at review time---to reconciliation, funding and institutional approval.
    Closing the file unlocks a \$60 reserve but freezes its allocations.
    \textbf{(b)} The agent must discover relevant documents and tools, obtain customer facts and request-specific approvals from the user simulator, and execute the linked workflows.
    Critical intermediate steps require correct reconciliation before closure and explicit consent to the revised \$60 total fee cap and the actual request terms before execution; final application approval alone cannot establish these requirements.
    Document labels report body word counts; amounts are USD and $t$ denotes logical minutes.}
    \label{fig:reconciliation-onboarding-case}
    \end{minipage}
\end{strip}

\section{A Complete Task Example}
\label{sec:case-dependencies}

Figure~\ref{fig:reconciliation-onboarding-case} introduces Q03, which links receipts reconciliation with an account application.
Panel~(a) presents the workflow dependencies, and panel~(b) illustrates the corresponding agent interaction.
We explain below how the funding decision connects these workflows.

The three routes in Table~\ref{tab:reconciliation-onboarding-draws} debit the same \$60 reserve after Acacia's file is closed.
In the Golden trace, the customer explicitly raises the total fee cap from \$50 to \$60 and approves route~B, which funds a single \$50 review submission at $t=8$; the application is approved at $t=18$, within the deadline of $t=19$.
From the same release time, route~A credits funds at $t=18$, leaving the ten-minute review beyond the deadline, while route~C falls one cent short of the submission fee.
These hypothetical comparisons are derived from the release terms.
The customer permits neither a top-up nor transfers from unrelated accounts.

\begin{table}[!htbp]
    \centering
    \captionsetup{justification=raggedright,singlelinecheck=false}
    \caption{Three release routes for the same \$60 reserve. The \$50 review fee is additional.}
    \label{tab:reconciliation-onboarding-draws}
    \setlength{\tabcolsep}{3pt}
    \renewcommand{\arraystretch}{1.08}
    \begin{tabularx}{\linewidth}{@{}>{\raggedright\arraybackslash}Xrrr@{}}
        \toprule
        \textbf{Release route} & \textbf{Net credit} & \textbf{Fee} & \textbf{Delay} \\
        \midrule
        A: Standard & \$60 & \$0 & 10 minutes \\
        B: Same-cycle & \$50 & \$10 & Immediate \\
        C: Reduced-net & \$49.99 & \$10.01 & Immediate \\
        \bottomrule
    \end{tabularx}
\end{table}

Each task comprises \emph{a customer objective and opening request}; \emph{a customer scenario} governing private facts and cooperation; \emph{an initial state} recording resources, histories, obligations, and scheduled events; \emph{access to shared documents and executable tools}; and \emph{deterministic acceptance criteria} for final outcomes and execution histories.
A replayable Golden trace and validation evidence establish feasibility and check the evaluator's treatment of valid alternatives and violations.
The displayed dialogue is constructed from a Golden trace; the SOP graph, Golden trace, and executable checks are not supplied to the evaluated agent.

\section{Dataset Statistics}
\label{sec:statistics_details}
\newcommand{\datasetstat}[1]{\csname datasetstat@#1\endcsname}
\newcommand{\setdatasetstat}[2]{\expandafter\gdef\csname datasetstat@#1\endcsname{#2}}
\setdatasetstat{task-count}{\numTasks{}}
\setdatasetstat{documents}{789}
\setdatasetstat{discoverable-tools}{601}
\setdatasetstat{permanent-agent-tools}{9}
\setdatasetstat{user-tools}{10}
\setdatasetstat{document-tokens}{598,827}
\setdatasetstat{document-length-mean}{759.0}
\setdatasetstat{document-length-median}{452}
\setdatasetstat{document-length-range}{44--10,755}
\setdatasetstat{base-collections}{398}
\setdatasetstat{nonempty-base-collections}{227}
\setdatasetstat{base-records}{2,994}
\setdatasetstat{reference-agent-calls-mean}{111.8}
\setdatasetstat{reference-agent-calls-median}{100.5}
\setdatasetstat{reference-agent-calls-range}{51--282}
\setdatasetstat{reference-user-calls-mean}{28.6}
\setdatasetstat{reference-user-calls-median}{21.5}
\setdatasetstat{reference-user-calls-range}{10--74}
\setdatasetstat{reference-documents-mean}{5.4}
\setdatasetstat{reference-documents-median}{5}
\setdatasetstat{reference-documents-range}{2--15}
\setdatasetstat{reference-records-mean}{29.9}
\setdatasetstat{reference-records-median}{23.5}
\setdatasetstat{reference-records-range}{4--96}
\setdatasetstat{instructions-mean}{14.2}
\setdatasetstat{instructions-median}{11}
\setdatasetstat{instructions-range}{6--36}

All statistics refer to the \datasetstat{task-count} released tasks. Golden trace statistics use one validated trace per task.
Table~\ref{tab:dataset_statistics} summarizes the tasks and shared environment.
Table~\ref{tab:dataset_distributions} reports document lengths and Golden trace statistics.

\begin{table*}[!t]
    \centering
    \captionsetup{justification=raggedright,singlelinecheck=false}
    \setlength{\tabcolsep}{3pt}
    \renewcommand{\arraystretch}{1.08}
    \begin{minipage}[t]{0.44\linewidth}
        \vspace{0pt}
        \caption{Task and environment counts in BizAgentBench.}
        \label{tab:dataset_statistics}
        \begin{tabularx}{\linewidth}{@{}>{\raggedright\arraybackslash}Xr@{}}
            \toprule
            \textbf{Statistic} & \textbf{Count} \\
            \midrule
            Tasks & \datasetstat{task-count} \\
            Always-available agent tools & \datasetstat{permanent-agent-tools} \\
            Discoverable business capabilities & \datasetstat{discoverable-tools} \\
            User-simulator tools & \datasetstat{user-tools} \\
            Knowledge documents & \datasetstat{documents} \\
            Total document-body tokens & \datasetstat{document-tokens} \\
            Queryable database collections & \datasetstat{base-collections} \\
            Database records & \datasetstat{base-records} \\
            \bottomrule
        \end{tabularx}
    \end{minipage}\hfill
    \begin{minipage}[t]{0.53\linewidth}
        \vspace{0pt}
        \caption{Document lengths and Golden trace statistics.}
        \label{tab:dataset_distributions}
        \begin{tabularx}{\linewidth}{@{}>{\raggedright\arraybackslash}Xrrr@{}}
            \toprule
            \textbf{Metric} & \textbf{Mean} & \textbf{Median} & \textbf{Range} \\
            \midrule
            \multicolumn{4}{@{}l}{\textit{Knowledge base: per document}} \\
            Document length (tokens) & \datasetstat{document-length-mean} & \datasetstat{document-length-median} & \datasetstat{document-length-range} \\
            \midrule
            \multicolumn{4}{@{}l}{\textit{Golden traces: per task}} \\
            Assistant tool calls & \datasetstat{reference-agent-calls-mean} & \datasetstat{reference-agent-calls-median} & \datasetstat{reference-agent-calls-range} \\
            User-simulator tool calls & \datasetstat{reference-user-calls-mean} & \datasetstat{reference-user-calls-median} & \datasetstat{reference-user-calls-range} \\
            Distinct documents read & \datasetstat{reference-documents-mean} & \datasetstat{reference-documents-median} & \datasetstat{reference-documents-range} \\
            Distinct business records read & \datasetstat{reference-records-mean} & \datasetstat{reference-records-median} & \datasetstat{reference-records-range} \\
            Executed business requests & \datasetstat{instructions-mean} & \datasetstat{instructions-median} & \datasetstat{instructions-range} \\
            \bottomrule
        \end{tabularx}
    \end{minipage}
\end{table*}

\noindent\textbf{Documents, tools, and records.}
Each document is counted once by its identifier.
Token counts use document bodies and the \texttt{cl100k\_base} tokenizer, excluding titles and metadata.
Tool counts distinguish always-available agent tools, discoverable business capabilities, and user-simulator tools across product portals.
Database counts describe the shared initial database before task-specific changes, including background records and empty collections; \datasetstat{nonempty-base-collections} of its \datasetstat{base-collections} queryable collections are nonempty.
Private customer files and hidden event payloads are excluded.
Collections are counted by query resource name.

\noindent\textbf{Golden traces.}
Assistant and user-simulator tool calls are counted separately, and batched calls count individually.
The traces average \datasetstat{reference-agent-calls-mean} assistant tool calls and \datasetstat{reference-user-calls-mean} user-simulator tool calls per task.

Golden traces demonstrate feasible strategies, not minimum-length solutions.

Each dispatch call counts once toward its target capability; unlocking a capability counts as a separate tool call.
Distinct documents are deduplicated by document identifier.
Distinct records count successful full-record reads, deduplicated by resource name and record identifier.
Executed business requests count distinct successfully executed request identifiers.

\section{Illustrative Business Cases}
\label{sec:business_cases}

The following examples focus on payment and delivery decisions in Q24 and Q34. All monetary amounts are in US dollars; times are logical minutes.

\subsection{A Limited Net-Delivery Guarantee (Q24)}
\label{case:net_guarantee}

Two suppliers must receive their exact remaining invoice balances under current banking instructions. The customer permits at most \$75 in total bank charges, including the earlier transfer's return charge. After that transfer resolves, the remaining balances are \$300 and \$400. A single-use guarantee offers exact net delivery for a fixed fee; the standard route exposes the recipient to intermediary deductions, followed by a separately charged top-up.

Consider assigning the guarantee to the \$300 balance and sending the \$400 balance through the standard route. Both suppliers ultimately receive the correct amounts. However, the \$20 return charge, \$20 guarantee fee, \$35 intermediary deduction, and \$5 top-up fee total \$80. The fee budget is exceeded despite correct final payments. The case requires allocating a limited service across obligations and accounting for charges incurred at different stages of settlement.

\subsection{Payment Reconciliation and Delivery Windows (Q34)}
\label{case:delivery_window}

A merchant must reconcile an already-paid order without charging again and deliver three compatible units: one rated at least~15 by minute~2, and two rated at least~10 by minute~5. All require the same socket. The priority delivery service takes one minute but must be booked by minute~0; ordinary delivery takes three minutes.

Consider repairing the merchant mapping and propagating two captures totaling \$10 and a \$1 refund, yielding the order's \$9 net payment, before waiting for the warehouse result at minute~1. The priority booking is then rejected because its window has closed. Compatible inventory is identified and the two later units arrive on time, but the early unit cannot meet its deadline through the remaining service. This sequence shows why an agent must coordinate information gathering with expiring service rights: correct reconciliation and suitable stock alone do not complete the commercial objective.

\section{Execution-History Analysis}
\label{sec:execution_analysis}

\subsection{Model execution length and tool use}
\label{sec:execution_length_details}

We analyze DeepSeek-V4.1-Flash execution traces in the default condition. Table~\ref{tab:deepseek_trace_distribution} counts batched calls individually and separates assistant and user-simulator tool calls. Database queries use \texttt{search\_records} or \texttt{get\_record}. An identical repeated database read has the same name, arguments, and response as an earlier read in the same run, excluding queries that observe changed state.

\begin{table}[!htbp]
    \centering
    \captionsetup{justification=raggedright,singlelinecheck=false}
    \caption{DeepSeek-V4.1-Flash tool use per run in the default condition.}
    \label{tab:deepseek_trace_distribution}
    \setlength{\tabcolsep}{3pt}
    \renewcommand{\arraystretch}{1.08}
    \begin{tabularx}{\linewidth}{@{}>{\raggedright\arraybackslash}Xrrr@{}}
        \toprule
        \textbf{Metric} & \textbf{Mean} & \textbf{Median} & \shortstack[r]{\textbf{90th}\\\textbf{percentile}} \\
        \midrule
        Assistant tool calls & 191.3 & 182 & 287 \\
        User-simulator tool calls & 28.2 & 26 & 41 \\
        Database queries & 111.1 & 101 & 180 \\
        Distinct documents read & 20.1 & 22 & 34 \\
        Identical repeated database reads & 16.4 & 14 & 32 \\
        \bottomrule
    \end{tabularx}
\end{table}

Successful runs average 13.7 identical repeated database reads, versus 18.5 in failed runs. Rechecking unchanged records is measurable redundancy, although some rechecks may be deliberate validation.

\subsection{User-simulator turn-taking reliability}
\label{sec:simulator_patch_ablation}

We examine whether the user simulator returns control to the agent after customer-side operations. After a successful write or wait, it must report the result before using another tool. We count batching multiple writes or waits, or continuing tool use before reporting the result and returning control, as protocol violations.

The comparison uses the same 50 tasks and DeepSeek-V4.1-Flash in both roles, reusing an existing complete evaluation for the revised prompt. The original prompt already requests one write at a time; the revision clarifies how to prioritize mixed questions and action requests and when to return control. Violations occur on four tasks (8\%) with the original prompt: two involve batched customer approvals and two involve continued portal use after a tool response. No such violations occur with the revised prompt (0\%; paired exact McNemar $p=0.125$). These observations motivate our use of the revised prompt in benchmark evaluations.

\subsection{Example of goal attainment with process violations}
\label{sec:goal_process_cases}

The single-run audit in the default condition contains two executions with $G=1$ and $P=0$.

\noindent\textbf{Rejected premature execution (Q13).}
The agent coordinates insurance settlement and a stored-value order payment. It invokes \texttt{execute\_request} on the payment before portal approval; the tool returns \texttt{CUSTOMER\_APPROVAL\_REQUIRED}, with no payment or balance change. The customer then approves and the agent completes the goals. The premature attempt violates approval discipline: $G=1$, $P=0$, $F=1$. No unauthorized transfer occurs.

\noindent\textbf{Actual reserve breach (Q18).}
The customer needs interest corrections and accountant payments: a \$60 retainer at minute~0 and a \$40 balance by minute~2, while retaining \$20 in the operating account throughout. The agent pays the retainer from the initial \$60, leaving zero. A \$18.66 refund settles and the agent posts a \$50 adjustment at minute~1; the final account balance after the remaining payment is \$28.66. The goals and closing reserve pass, but the earlier zero balance violates the continuous requirement: $G=1$, $P=0$, $F=1$. Only the execution history exposes this breach.

\subsection{Business-document scope in retrieval}
\label{sec:business_document_scope_example}

In Q01, the customer asks the agent to pay suppliers in euros and pounds and match earlier tax payments to the correct periods. The applicable Operations Desk rules are described in \texttt{DESK-tax} and \texttt{DESK-fx}. Searching for \texttt{tax payment matching periods reconciliation} returns \texttt{PX-B-77}, an overpayment-credit document for a different product, in first position and \texttt{PX-B-76}, a tax-agent document for that product, in fifth position. The agent opens both documents, whose terminology matches the query although neither governs this Operations Desk task.

\end{document}